# A New Fast Unweighted All-pairs Shortest Path Search Algorithm Based on Pruning by Shortest Path Trees

Yasuo Yamane[1] and Kenichi Kobayashi[2]

[1] Fujitsu Limited, 17-25, Shinkamata 1-chome, Ota-ku, Tokyo, 144-8588, Japan

[2] Fujitsu Laboratories Ltd., 4-1-1, Kamikodanaka, Nakahara-ku, Kawasaki, 211-8588, Japan

Abstract

We present a new fast all-pairs shortest path algorithm for unweighted graphs. In breadth-first search which is said to representative and fast in unweighted graphs, the average number of accesses to adjacent vertices (expressed by α) is about equal to the average degree of the graph. On the other hand, our algorithm utilizes the shortest path trees of adjacent vertices of each source vertex, and reduce α drastically in comparison with breadth-first search. Roughly speaking α is reduced to the value close to 1, because the average degree of a tree is about 2, and one is used to come in and the other is used to go out, although that does not hold true when the depth of the shortest path trees is small. We compared our algorithm with breadth-first search algorithm, and our results showed that ours outperforms breadth-first search on speed and α.

1. Introduction

We present a new fast all-pairs shortest path algorithm for unweighted graphs, which we call PST (Pruning by Shortest path Trees) algorithm, or PST simply below.

All pairs shortest path algorithms have many applications in general graphs, for example, railroad networks, transportation networks, Web and SNS (Social Networking Service). Restricted to unweighted graphs, they have various applications such as Web and SNS.

Breadth-first search is applied for various purposes, and it can be applied for computing the all-pairs shortest paths [BFS]. We call it simply BFS. It is one of the representative algorithms for unweighted all-pairs shortest path search and known to be fast. In BFS, the average number of accesses to adjacent vertices at each vertex, which is expressed by $\alpha$ below, is about equal to the average degree of the graph. On the other hand, PST utilizes the shortest path trees of adjacent vertices of each source (starting) vertex for pruning when traversing from the source vertex beyond the adjacent vertices and reduce α drastically in comparison with BFS. And roughly speaking $\alpha$ of PST is reduced to the value close to 1, because the average degree of a tree is about 2, and one is used to come in and the other is used to go out, although that does not hold true when the depth of the shortest path trees is small.

We compared PST with BFS, and our result showed that PST outperforms BFS on speed and $\alpha$.

2. Relative Works

In this section, we explain representative algorithms for all-pairs shortest path algorithms for unweighted graphs as follows:

1) Floyd-Warshall algorithm ([Floyd62], [Warshall62])

It is one of the most famous algorithms for all-pairs shortest algorithms. This is the algorithm for weighted graphs, but it can be applied to unweighted graphs letting all weights be one. The (worst) time complexity is $O(n^3)$ where $n$ is the number of vertices, but the implementation is very simple requiring only some lines. Let $n$ and $m$ mean the numbers of vertices and edges of a graph respectively, and time complexity mean worst one below.

2) Dijkstra's algorithm ([Dijkstra])

It is well-known as a fast algorithm for computing the shortest paths from a source vertex to the all vertices in weighted graphs, and the time complexity is $O(n \log n + m)$. It can be applied for computing the all-pairs shortest paths by letting each vertex as a source one, and the time complexity is $O(n(n \log n + m))$. To distinguish them, we call the former algorithm SS-Dijkstra (Single-Source Dijkstra), and the latter AP-Dijkstra (All-Pairs Dijkstra) or simply Dijkstra.

3) breadth-first search ([BFS])

As mentioned above, breadth-first search is originally an algorithm to traverse all the vertices in breadth-first manner, and it is applied for various purposes. The time complexity is $O(n + m)$. It can be applied for computing the all-pairs shortest paths by letting each vertex as a source one, and the time complexity is $O(n(n + m))$. To distinguish from an original breadth-first search, we call this algorithm AP-BFS (All-Pairs shortest path Breadth-First Search) or simply BFS. AP-BFS is faster than AP-Dijkstra and Floyd-Warshall algorithm on unweighted graphs.

4) Peng's algorithm ([Peng12])

This is a variant of AP-Dijkstra's algorithm, and it computes the shortest paths from each vertex to all the vertices serially. We call this algorithm Peng simply. The feature of it is to utilize the length of the already computed shortest paths to reduce $\alpha$. AP-BFS is faster than AP-Dijkstra, so we thought of improving Peng for unweighted graphs likewise, but we have not achieved that. We will submit a paper on a variant of PST for weighted graphs to arXiv at almost the same time as this paper, so we will compare PST with Peng in the paper.

3. PST algorithm

In this section, PST algorithm is explained in more detail. Our basic idea is explained in 3.1 and the details of the algorithm is explained in 3.2.

3. 1 Basic idea

Generally speaking, in all-pairs shortest path algorithms, a shortest path tree is generated at each vertex. It is well-known that the shortest paths from each vertex $v$ to all the vertices can be compactly expressed by a tree whose root is $v$. This tree is called a "shortest path tree" and represented by $T(v)$ below. Fig.3.1(b) shows the shortest path tree for an unweighted graph shown in (a).

As mentioned above, PST utilizes the shortest path trees being generated for pruning, that is, to reduce α. Here, let us consider how similar $T(v)$ and $T(w)$ are if $v$ and $w$ are adjacent; for example, your closest railway station $v$ and another station $w$ next to $v$. $T(v)$ and $T(w)$ must be almost the same. This suggests that $T(w)$ has only to be traversed, that is, it is unnecessary to traverse the edges which are not contained in $T(w)$ when searching through $w$ to generate $T(v)$ if the necessary part of $T(w)$ is already generated. This leads to reducing α drastically in comparison with BFS, as mentioned above. This is our basic idea.

PST is a variant of BFS where the following two modifications are added. The shortest path from $v$ to $x$ on $T(v)$ is expressed by $\sigma_v(x)$.

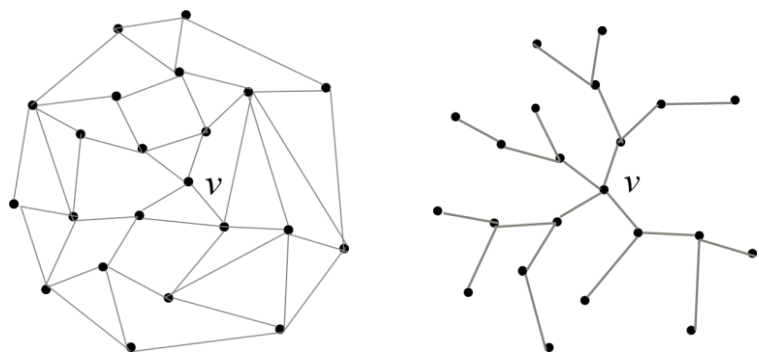

(a) an unweighted graph    (b) a shortest path tree

Fig.3.1 An unweighted graph and a shortest path tree

1) Pruning by shortest path trees

Let us consider two adjacent vertices $v$ and $w$, and another vertex $x$ different from them. It is well known that a partial path of a shortest path is also the shortest. Therefore, if $\sigma_v(x)$ contains edge $(v,w)$, then $\sigma_v(x)$ can be let to be edge $(v,w)$ plus $\sigma_w(x)$. This means that when searching through $w$ from $v$, it is unnecessary to search the part other than $T(w)$. Let the adjacent vertices of $v$ be $w_1, w_2$ and $w_3$. Fig.3.2 shows the parts of the three shortest path trees $T(w_1), T(w_2)$ and $T(w_3)$ which are traversed to create $T(v)$. They are expressed as red, green, and blue edges respectively.

Based on what is mentioned above, PST traverses only $T(w)$ when searching through $(v,w)$. As mentioned above, this improvement reduces $\alpha$ drastically in comparison with BFS. Roughly speaking to make the effect understandable, $\alpha$ is reduced to the value close to one. Because the average degree of a tree is about 2 (correctly $\frac{2(n-1)}{n}$), and on each vertex $x$, one of the adjacent vertices of $x$ is used to go into it and the other is used to access the adjacent vertex, so $\alpha$ is about 1. However, this does not hold true when the depth of shortest path trees is small. Here, it seems worth to note the accesses to adjacent vertices from a source vertex and those from adjacent vertices to their adjacent vertices cannot be reduced to understand that it does not hold true.

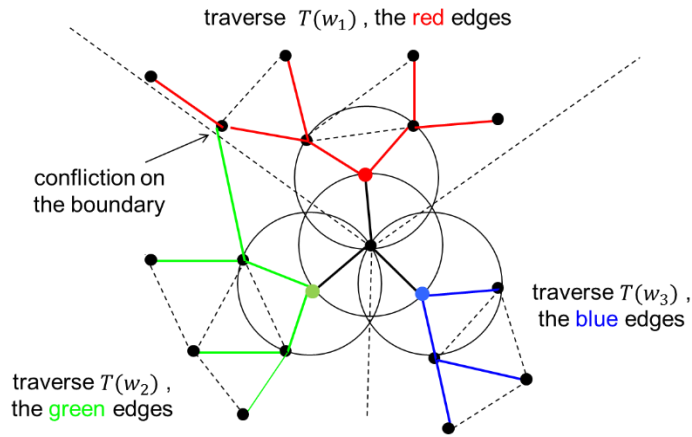

Fig.3.2 Traversing the shortest path trees of adjacent vertices to create $T(v)$

2) Synchronous generation of shortest path trees

In PST it is necessary to generate shortest path trees synchronously to realize 1). Concretely speaking, at each vertex $v$, the parts of $T(v)$ is synchronously generated, and they are like concentric circles. First the part of the vertices adjacent to $v$, that is, the vertices on the circle centering around $v$ whose radius is 1 is generated synchronously with all the other vertices. Secondly the part of the vertices on the circle whose radius is 2 is generated synchronously as if the computation is done parallelly, and so on. Therefore, letting a vertex $w$ is adjacent to $v$, when generating $T(v)$ it is guaranteed that the necessary part of $T(w)$ has already be generated.

3.2 The details of PST algorithm

Here, we explain concretely how the modifications mentioned above are added. We implemented for undirected graphs but it is easy to modify for directed graphs. We add notice when necessary.

1) Modification for pruning by shortest path trees

This is modification of the data structure representing shortest path trees. In BFS, $T(v)$ of each source vertex $v$ can be represented as a vector of size n. That is, letting all the vertices be $v_1, v_2, \cdots, v_n$ and the vector be $[s_i]$, it is sufficient to let $s_i$ be $j$ of the parent $v_j$ of $v_i$ on $T(v)$. And all the shortest path trees can be represented by an $n$ by $n$ matrix. However, in PST, it is necessary to memorize children in shortest path trees to prune using shortest path trees. So, for each source vertex, a shortest path tree retaining the relationship between the parents and the children is generated.

2) Modification for synchronous generation of shortest path trees

This is modification on queues. PST uses FIFO queues in the same manner as BFS, but it is necessary to modify them to synchronize as follows: A pair of a vertex and its distance from the source vertex is enqueued. And only the pair at the top of the queue whose distance is equal to a specified distance is dequeued. This modification enables to add synchronously the vertices to the shortest path tree whose distance from the source vertex is the same. We call this queue a "queue with distances" or a "d-queue" simply. Although we do not explain the class of d-queue in detail, each d-queue $q$ has two methods. One is $q.enque(x, d)$ which is used to enqueue a vertex $x$ whose distance from $v$ is $d$, and the other is $q.deque(d)$ which returns $x$ whose distance from $v$ is $d$ at the top of the queue, or $None$ when such a vertex does not exist. We implemented d-queue as a circular list of size $s$. Likewise, we implemented the FIFO queue in BFS as a circular list of size $s$.

We show the two classes: One is $Vertex$, and its instance represents a vertex of a graph $G$. The other is $T\_Vertex$ and its instance represents a vertex on a shortest path tree, which is called "t-vertex" to distinguish from a vertex of $Vertex$.

Let $x$ be a vertex. Then we express a t-vertex whose id is the same as $x$ by $x'$. Then we say $x'$ corresponds to $x$.

1) $Vertex$

Let an instance of this class be $v$. It consists of five properties $id$, $adj\_veritces$, $root$, $c$ and $que$. Each vertex $v_i$ ($1 \le i \le n$) has an id whose value is $i$, and property $id$ represents $i$. Property $adj\_veritces$ represents the set of adjacent vertices of $v$ whose class is also $Vertex$. Property $root$ represents a vertex $v'$ in $T(v)$ corresponding to $v$, that is, the root of $T(v)$. The class of $v'$ is $T\_Vertex$ explained next. Property $c$ represents a counter for vertices whose shortest paths from $v$ have already been obtained. Property $que$ represents a d-queue prepared for each vertex explained above.

| | |
|---|---|
| $id$ | the id of $v$ from 1 to $n$ |
| $adj\_veritces$ | the set of adjacent vertices of $v$ whose class is $Vertex$ |
| $root$ | the root of $T(v)$ whose class is $T\_Vertex$ |
| c | a counter for vertices whose shortest paths have already been obtained. |
| $que$ | a d-queue for $v$ |

2) $T\_Vertex$

As mentioned above, an instance of this class is used to represent each vertex of $T(v)$. $T$ of $T\_Vertex$ means (shortest path) tree. Let $x'$ be a t-vertex of $T(v)$. The t-vertex $x'$ has four properties $vertex$, $cor$, $parent$, and $children$. Property $vertex$ represents $x$. Here, let $x'$ be a t-vertex reached through $w'$ where $w$ and a source vertex $v$ are adjacent. Property $cor$ represents the t-vertex corresponding to $x'$ on $T(w)$. This t-vertex is represented by $x''$ below. That is, the value of $cor$ is $x''$. Property $parent$ represents the parent of $x'$ in $T(v)$. Property $children$ represents the set of the children of $x'$ in $T(v)$.

The values of properties of $x'$ and the relationships between these data structures are summarized in Fig.3.3.

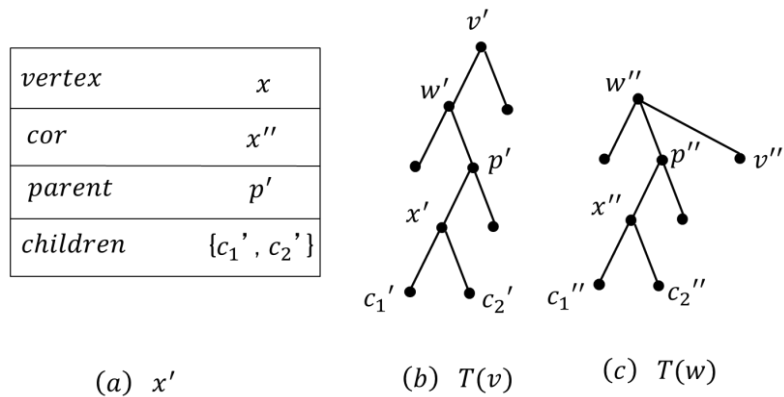

Fig3.3 the data structure of a t-vertex and the relationship between t-vertices and shortest path trees

The python-like pseudo code of PST algorithm is shown below. First the algorithm for creating a graph $G = (V, E)$ where $|V| = n$ and initialization is shown. We implemented the creation of a graph and initialization together in our experiments, but they should be separated to be exact because it relates to the measurements as mentioned later. Here, initialization means preparation for PSTw algorithm; for example, initialization of a queue at each vertex.

In python's class definition, the name of a method creating a new instance is $\_init\_$, and the first parameter of it is $self$ which represents a new instance. However, we use an understandable parameter name instead of $self$, for example $v$.

---

Algorithm for creation of a graph and initialization

   input: $n$

   output: $G = (V, E)$

---

class $Vertex()$:

  $\_init\_(v, id, n)$:      # returns a new instance $v$

    $v.id = id$

    $v.adj\_vertices = \{\}$

    $v' = T\_Vertex(v)$

    $v.root = v'$
    $v.c = 1$
    set an empty d-queue of size $n$ to $v.que$

$class\ T\_Vertex():$
    $\_init\_(v', v):$          # returns a new instance $v'$
    $v'.vertex = v$
    $v'.cor = None$
    $v'.parent = None$
    set an empty set to $v'.children$

set an empty set to $V$
$for\ id \in \{1, 2, \cdots, n\}:$
    $v = Vertex(id, n)$     # create a new instance of $Vertex$
    add $v$ to $V$
$for\ v \in V:$
    add $v's$ adjacent vertices to $v.adj\_veritces$

Secondly the main routine of PST algorithm, and the definition of function $extend(v, d, D, S)$, which extends $T(v)$, are shown. A variable $d$ represents the depth of shortest path trees being generated. $D$ is an $n$ by $n$ matrix to represent distances between vertices. It is called a "distance matrix". Its element $D[i,j]$ means the distance from $v_i$ to $v_j$. All the elements of $D$ are initialized to 0.0. $S$ is an $n$ by $n$ matrix to represent all the generated shortest path trees. It is called a "shortest path tree matrix". Its element $S[i,j]$ represents the parent of $v_i$ on the shortest path from $v_i$ to a source vertex $v_j$. Let the diagonal elements of $S$ already be initialized to $NO\_PARENET\,(= -1)$, and all the other elements to $NOT\_SEARCHED\,(= 0)$. $NO\_PARENET$ means there is not a parent, that is, $v_i$ is a source vertex. $NOT\_SEARCHED$ means it has not been searched yet.

The function returns if $v.c == n$; BFS was implemented likewise. It is necessary to note that this condition cannot be used for directed graphs because all the vertices cannot necessarily be reached from a source vertex.

It is worth noting that although at first $V$ represents the set of all the vertices, $V$ virtually represents the set of vertices whose all the shortest paths form source vertices have not been computed yet.

PST algorithm
    input: $G = (V, E)$

output: $D$ and $S$

---

set an $n \times n$ initialized distance matrix to $D$
set an $n \times n$ initialized shortest path tree matrix to $S$
$d = 0$
while $0 < |V|$:
    $d = d + 1$
    set an empty set to $V_{new}$
    for $v \in V$:
        $extend(v, d, D, S)$:
        if $v.c < n$:
            add $v$ to $V_{new}$
    $V = V_{new}$

$def\ extend(v, d, D, S)$:
    if $d == 1$:
        for $w \in v.adj\_vertices$:
            $D[w.id, v.id] = d$
            $S[w.id, v.id] = v.id$
            $w' = T\_Vertex(w)$       # create a new instance of $T\_Vertex$
            $w'.cor = w.root$       # the value of $w.root$ is $w''$
            add $w'$ to $v'.children$
            $w'.parent = v'$
            $v.que.enque(w', 1)$
            $v.c = v.c + 1$
    else:
        $n = len(D)$
        $w' = v.que.deque(d - 1)$
        while $w'$ is not None:
            for $x'' \in w'.cor.children$:
                $x = x''.vertex$
                if $S[x.id, v.id] == NOT\_SEARCHED$:
                    $D[x.id, v.id] = d$
                    $w = w'.vertex$
                    $S[x.id, v.id] = w.id$
                    $x' = T\_Vertex(x)$       # create a new instance of $T\_Vertex$
                    $x'.cor = x''$

```
            add x' to w'.children
            x'.parent = w'
            v.que.enque(x', d)
            v.c = v.c + 1
            if v.c == n: return
        w' = v.que.deque(d − 1)
return
```

4. Evaluation

We compared PST with BFS. We excluded Floyd-Warshall algorithm (abbreviated as WF) because its time complexity is $O(n^3)$ which is worse than $O(n(n+m))$ of BFS. We measured letting $n = 2^{2i} (i = 3,4,5,6)$, that is, $n = 64, 256, 1024, 4096$.

On CPU time, as mentioned above, creating a graph also includes initialization. Therefore, CPU time of PSTw does not include the initialization time. We think initialization does not affect the total CPU time so much, but it should be included in PSTw to be exact. Peng and Dijkstra are also implemented in the same manner, so their CPU time also do not include initialization time.

We used the following two kinds of graphs for comparison, which are hypercube-shaped and scale-free graphs. We selected scale-free graphs because they are said to be ubiquitous in the real world. The relationship between the degree of each vertex $d$ and the frequency of the vertices whose degree is equal to $d$ obey to a power distribution. So $d$ takes various values. On the other hand, hypercube-shaped graphs have a feature quite opposite to scale-free graphs, that is, $d$ takes only one value. We selected hypercube-shaped graphs because we wanted to examine the two algorithms from two quite different viewpoints.

1) hypercube-shaped graph

It is worth to note that the degree of each vertex of this graph is $\log_2 n$, the values of $\alpha$ are about 6, 8, 10, and 12 for $n = 64, 256, 1024$, and $4096$.

2) Scale-free graph

The graph is created as follows: When creating a graph of size $n$, first a complete graph $G = (V, E)$ of size $n'(< n)$ is created. Secondly the remaining $n - n'$ vertices are added one by one as follows: Let one of them be $v$. Let $n'$ vertices chosen randomly from $V$ be $v_1, v_2, \cdots, v_{n'}$. Then Let $V = V \cup \{v\}$ and $E = E \cup \{(v, v_1), (v, v_2), \cdots, (v, v_{n'})\}$. The probability of choosing $v_i$ $(i = 1, 2, \cdots, n')$ is let to be proportional to the degree of $v_i$.

For measurement environment, we used FUJITSU Workstation CELSIUS M740 with Intel Xeon

E5-1603 v4 (2.80GHz) and 32GB main memory, programing language Python, and OS Linux.

4.1 Comparison in hypercube-shaped graphs

Fig. 4.1 and Table 4.1 show the comparison of CPU time of the two algorithms in hypercube-shaped graphs. Fig. 4.1 is shown in a double-logarithmic graph. CPU time graphs are shown in the same manner below. Table 4.1 shows the actual values in detail. Tables show the actual values likewise below. The column whose name is /PST shows the rate of the CPU time (called "CPU time rate") of BFS against that of PST, that is, how many times PST is faster than BFS. When $n = 4096$, PST is 3.1 times faster than BFS, and the CPU rate increases as $n$ increases.

Fig. 4.2 and Table 4.2 show the comparison of $\alpha$. The /PST column shows the rate of $\alpha$ (called "$\alpha$ rate") likewise. PST outperforms BFS greatly, and the $\alpha$ rate increases as $n$ increases. BFS's $\alpha$ is about equal to the degree of each vertex and increases as $n$ increases. PST's does not reach 1 but decreases from 1.71 to 1.52, a close value to 1.

To sum up, PST outperforms BFS in CPU time and $\alpha$, and especially greatly in $\alpha$. The $\alpha$ rate is larger than CPU rate. It seems to be because of the overhead of t-vertices.

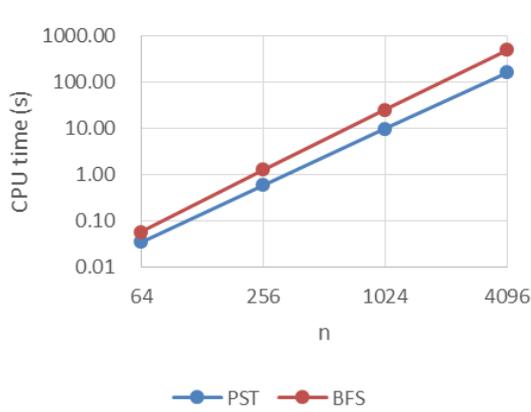
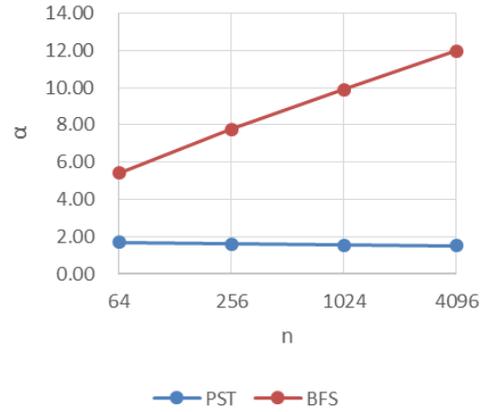

Fig. 4.1 Comparison in CPU time　　　　　　Fig. 4.2 Comparison in α

Table 4.1 Comparison in CPU time

| n | PST time (s) | BFS time (s) | /.PSTw |
|---|---|---|---|
| 64 | 0.04 | 0.06 | 1.64 |
| 256 | 0.62 | 1.31 | 2.12 |
| 1024 | 10.17 | 25.61 | 2.52 |
| 4096 | 163.73 | 504.20 | 3.08 |

Table 4.2 Comparison in α

| n | PST α | BFS α | /.PSTw |
|---|---|---|---|
| 64 | 1.71 | 5.42 | 3.17 |
| 256 | 1.60 | 7.75 | 4.84 |
| 1024 | 1.54 | 9.90 | 6.43 |
| 4096 | 1.52 | 11.96 | 7.87 |

4.2 Comparison in scale-free graphs

The result of comparison in sparse case ($n' = 2$) is mentioned in 4.2.1, and that in dense case ($n' = \sqrt{n}$) is mentioned in 4.2.2.

4.2.1 Sparse case ($n' = 2$)

Fig. 4.3 and Table 4.3 shows the comparison of the two algorithms in CPU time. These show PST outperforms BFS slightly. When $n = 4096$, PST is 1.38 times faster than BFS and the CPU time rate increases as $n$ increases.

Fig.4.4 and Table 4.4 shows the comparison in $\alpha$. When $n = 4096$, PST's $\alpha$ is 1.19, a value to close to 1, and PST outperforms BFS by 3.26 times. The $\alpha$ rate increases as $n$ increases.

To sum up, PST outperforms BFS, but not so much as in case of hypercube-shaped graphs.

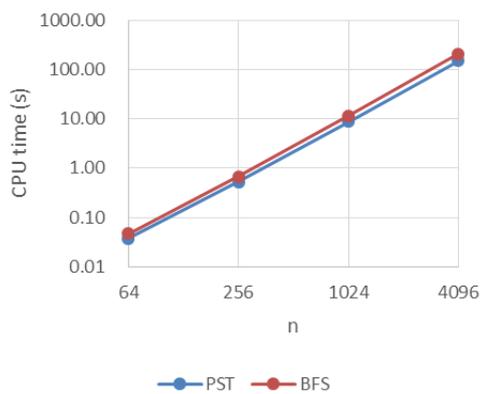
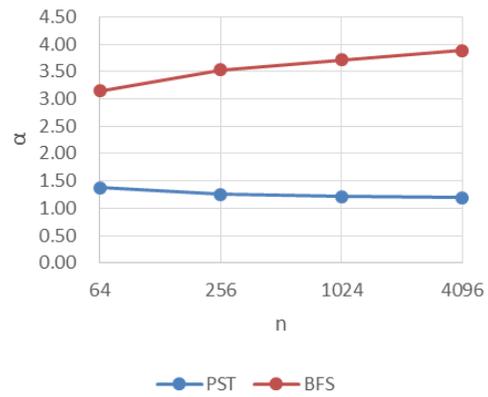

Fig. 4.3 Comparison in CPU time　　　　　　Fig. 4.4 Comparison in α

Table 4.3 Comparison in CPU time

| n | PST time (s) | BFS time (s) | /.PSTw |
|---|---|---|---|
| 64 | 0.04 | 0.05 | 1.24 |
| 256 | 0.53 | 0.68 | 1.28 |
| 1024 | 8.86 | 11.63 | 1.31 |
| 4096 | 148.94 | 206.04 | 1.38 |

Table 4.4 Comparison in α

| n | PST α | BFS α | /.PSTw |
|---|---|---|---|
| 64 | 1.38 | 3.15 | 2.28 |
| 256 | 1.25 | 3.53 | 2.82 |
| 1024 | 1.21 | 3.71 | 3.07 |
| 4096 | 1.19 | 3.88 | 3.26 |

4.2.2 Dense case ($n' = \sqrt{n}$)

Fig.4.5 and Table 4.5 shows the comparison of CPU time. PST is 1.42 times faster than BFS when $n = 4096$, and the CPU time rate increases as $n$ increases.

Fig. 4.6 and Table 4.6 show the comparison of α. PST outperforms BFS by 1.95 times when $n = 4096$, and the α rate seems to increase.

PST's α is 6.23 when $n = 4096$, and increases as $n$ increases. The depth of shortest path trees tends to be small when the graph is dense. As mentioned above, it does not holds true when the depth of shortest path tree is small that the value of α is close to 1, but it is smaller than in case of BFS.

To sum up, PST outperforms BFS on CPU time as well as in the sparse case, but not so much on α as in the sparse case. It does not holds true that the value of α is close to 1,

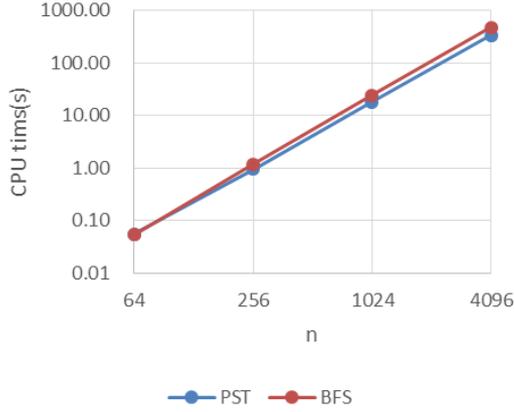
Fig. 4.5 Comparison in CPU time

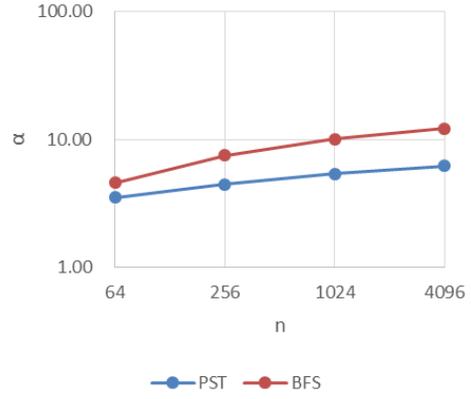
Fig. 4.5 Comparison in α

Table 4.6 Comparison in CPU time

| n | PST | BFS | |
|---|---|---|---|
| | time (s) | time (s) | /.PSTw |
| 64 | 0.06 | 0.05 | 0.98 |
| 256 | 0.93 | 1.17 | 1.25 |
| 1024 | 17.88 | 24.05 | 1.35 |
| 4096 | 333.72 | 472.44 | 1.42 |

Table 4.6 Comparison in α

| n | PST | BFS | |
|---|---|---|---|
| | α | α | /.PSTw |
| 64 | 3.50 | 4.58 | 1.31 |
| 256 | 4.47 | 7.52 | 1.68 |
| 1024 | 5.36 | 10.02 | 1.87 |
| 4096 | 6.23 | 12.12 | 1.95 |

4.3 Consideration

We show our consideration on the result of comparison and space complexity.

4.3.1 On the result of comparison

1) Comparison with BFS

PST outperforms BFS in CPU time and $\alpha$ in all the graphs measured. The reason seems as follows: BFS's $\alpha$ is about equal to the average degree of each vertex. On the other hand, BFS's $\alpha$ is close to 1 in case of hypercube-shaped and sparse scale-free graphs. In case of dense scale-free graphs, BFS's $\alpha$ is 6.23 and not close to 1, but it is 1.95 times smaller than BFS's $\alpha$ 12.1. In short, PST's $\alpha$ is at least 1.95 times smaller than BFS's $\alpha$, and that seems to be the reason why PST outperforms BFS on CPU time.

2) PST's $\alpha$

It is between 1.52 and 1.71 in hypercube-shaped graphs and it decrease as $n$ increases, and it is also between 1.19 and 1.38 in sparse scale-free graphs and it also decreases as $n$ increases and these values are closes to 1. On the other hand, in case of dense scale-free graphs it is between 3.50 and 6.23, which are far from 1. The reason seems as follows: Let vertex $w$ be adjacent to a source vertex $v$. Then $T(w)$ contains all the vertices adjacent to $w$, so $T(v)$ cannot be pruned at the depth of 1. In addition, the depth of $T(v)$ is small. These seem to cause that $\alpha$ is far from 1. The value of $\alpha$ increases as $n$ increases.

4.3.2 On space complexity

We mention our consideration about space complexity, which has not been discussed above. Both PST and BFS needs $O(n^2)$ memory to store the matrix for shortest path trees. In addition, BFS only needs $O(n + m)$ memory to represent a graph, and memory for the FIFO queue. However, in addition to the memory mentioned above, PST needs the memory to store $T(v)$ at each vertex $v$. Therefore, BFS outperforms PST from the viewpoint of space complexity.

5. Conclusion

We proposed a new all-pairs shortest path search algorithm for unweighted graphs, which reduces $\alpha$ to the value close to 1 by pruning based on the shortest path trees of adjacent vertices of a source vertex when the depth of the shortest paths is relatively large.

The result mentioned above showed the following:

1) PST outperforms BFS in CPU time and $\alpha$.
2) PST's $\alpha$ is close to 1 in case of hypercube-shaped and sparse scale-free graphs, and it decreases as $n$ increases
3) In case of dense scale-free graphs, PST's $\alpha$ is between 3.50 and 6.23 and far from 1.
   The reason seems that the shortest paths cannot be pruned at the depth of 1 and the depth of them is small.
4) BFS outperforms PST from the viewpoint of space complexity.